\documentclass[11pt]{article}
\def\ph{\phi}

\usepackage{graphicx}
\usepackage{placeins}
\usepackage{float}
\usepackage{subfigure}
\usepackage[toc,page]{appendix}
\usepackage{braket}
\usepackage{color,soul}
\usepackage{amsmath,amssymb}
\usepackage{epsfig}
\begin{document}
%%%%%%%%%%%%%%%%%%%%%%%%%%%%%%%%%
%%%%%%%%%%%%%%%%%%%%
\title{{\bf{\Large Gravitational surface Hamiltonian and entropy quantization}}}
%%%%%%%%%%%%%%%%%%%%
\author{ 
{\bf {\normalsize Ashish Bakshi$^{a}$}$
$\thanks{E-mail: ashishbakshi@outlook.com}}\,\ , \,\
{\bf {\normalsize Bibhas Ranjan Majhi$^{b}$}$
$\thanks{E-mail: bibhas.majhi@iitg.ernet.in}}  \,\ and \,\
 {\bf {\normalsize Saurav Samanta$^{c}$}$
$\thanks{E-mail: srvsmnt@gmail.com}} \\\\
{\normalsize $^{a}$Indian Statistical Institute,}
\\{\normalsize 203 B.T. Road, Kolkata-700108, India} \\\
{\normalsize $^{b}$Department of Physics, Indian Institute of Technology Guwahati,}
\\{\normalsize Guwahati 781039, Assam, India}\\\
{\normalsize $^{c}$Narasinha Dutt College,}
\\{\normalsize 129, Belilious Road, Howrah-711101, India}
\\[0.3cm]
}
%\date{}

\maketitle

\begin{abstract}
   The surface Hamiltonian corresponding to the surface part of a gravitational action has $xp$ structure where $p$ is conjugate momentum of $x$.  Moreover, it leads to $TS$ on the horizon of a black hole. Here $T$ and $S$ are temperature and entropy of the horizon. Imposing the hermiticity condition we quantize this Hamiltonian. This leads to an equidistant spectrum of its eigenvalues. Using this we show that the entropy of the horizon is quantized. This analysis holds for any order of Lanczos-Lovelock gravity. For general relativity, the area spectrum is consistent with Bekenstein's observation. This provides a more robust confirmation of this earlier result as the calculation is based on the direct quantization of the Hamiltonian in the sense of usual quantum mechanics.
\end{abstract}
\section{\label{Intro}Introduction}
   Quantum mechanically, black holes are thermodynamic objects which have temperature \cite{Hawking:1974sw} as well as entropy \cite{Bekenstein:1973ur}. The exact expressions for these entities have been obtained by semi-classical treatment. In the absence of a true ``quantum gravity'' theory, the precise microscopic origin of entropy is unknown. Since there is no such complete theory,{\footnote{Of course, there are some attempts like loop quantum gravity \cite{Ashtekar:1997yu}, string theory \cite{Strominger:1996sh}. But none of them are self sufficient.}} we can only do semi-classical computations. One of the earlier attempts to describe these microstates, which are responsible for horizon entropy, originated from the seminal works of Bekenstein \cite{Bekenstein:1973ur}. He showed that when a neutral particle is swallowed by a Kerr black hole, the lower bound on the increment of the horizon area is 
   \begin{equation}
   (\Delta A)_{\textrm {min}} = 8\pi l_p^2~,
   \label{Bekenstein}
   \end{equation}
   when the particle has a finite size -- not smaller than Compton wavelength. Here $l_p$ is the Planck length. After Bekenstein, the same was calculated for the assimilation of a charged particle \cite{Hod:1999nb} and it was found to have a similar Planck length square nature, but with a different numerical factor.
   
   Even now, people are still trying to understand and find the implications of this result. One of the most important consequences is -- it leads to the quantization of horizon area. A much more sophisticated calculation by Bekenstein and Mukhanov \cite{Bekenstein:1995ju} yielded the area spectrum as $4l_p^2\ln k$ with $k=2$. An exactly identical expression was later derived by using quasinormal modes, except with $k=3$ \cite{Hod:1998vk,Kunstatter:2002pj}. The importance of this result are as follows. It was found that this is consistent with the Gibbs' paradox \cite{Kiefer:2007uw} and also that the value of Immirzi parameter, in the context of Loop quantum gravity, can be fixed \cite{Dreyer:2002vy}. But unfortunately the derivation by quasinormal modes encountered a problem. Maggiore \cite{Maggiore:2007nq} observed that the imaginary part of the ringing frequency dominates compared to the real part in higher $n$ limit where $n$ is an integer. Now since the whole calculation is semi-classical, which is reliable at large $n$, one should take the imaginary part in the computation. Then this leads to the old result by Bekenstein: quantum of area is $8\pi l_p^2$. Consequently, several attempts \cite{Setare:2003bd}--\cite{Myung:2010af} have been made to find the spectrum of area or entropy. It turned out that the quantum of entropy is much more natural than that of area \cite{Banerjee:2009pf,Kothawala:2008in}. In all cases, for Einstein's gravity, one finds that the spacing is given by (\ref{Bekenstein}).
   
   Now it is quite evident that the semi-classical calculation is mostly in favour of Eq. (\ref{Bekenstein}). In this paper we make another attempt to quantize black hole entropy and area. The basic idea follows from an earlier result by one of the authors \cite{Parattu:2013gwa}. It has been observed that the surface part of the Einstein-Hilbert action has a structure like $xp$ where $x$ and $p$ are coordinate and conjugate momentum, respectively. Also the evaluation of the surface term on the horizon leads to the surface Hamiltonian which is the product of entropy and temperature. Moreover, one can identify that the entropy is equivalent to $x$ while the temperature plays the role of conjugate momentum $p$ (For details, see \cite{Parattu:2013gwa}). So everything reduces to a classical Hamiltonian $H=xp$. An extension to Lanczos-Lovelock gravity theory has also been done \cite{Chakraborty:2014joa} and here also same conclusion was drawn.
   
  % Hamiltonian of this form, though atypical in mechanics, is extremely important in literature. Its significance was first seen in the context of a long-standing number theory problem known as Riemann hypothesis. One attempt to prove this famous hypothesis is to find a self adjoint linear operator whose eigenvalues are the Riemann zeros. This was suggested long ago by Hilbert and Polya.
   
 %  Connes \cite{connes} took the the Hamiltonian $xp$ and used mathematical structures adeles (p-adic numbers) and ideles (associated units) in an attempt to solve Riemann hypothesis. Finding inspiration from this work, Berry and Keating \cite{berry1,berry2} made some analysis on this Hamiltonian to connect this with Riemann zeros. Though these two approaches are quite different in nature, the first more abstract and the second more intuitive, perhaps they are equivalent \cite{berry1}. In fact, both of these analyses are semiclassical and counting of states are done with different regularizations imposed on the phase space variables. Later, a similar study was done with regularization only on the position variable \cite{sierra}. All these regularizations \cite{Sierra:2007du}, though lack physical motivation, serve the purpose of energy quantization. This point is important, because the classical Hamiltonian $xp$ leads to unbounded solutions.
   
   In the present work, we discuss the quantization of $xp$ Hamiltonian. The standard quantization procedure leads to some confusions. The key point is that, the quantum Hamiltonian $\hat{H}=\hat{x}\hat{p}$ is not hermitian. So one might think that, a simple symmetric ordering of $\hat{H}$ will solve the problem. Unfortunately, there are few subtle points in this apparently simple program. As we shall see, self-adjoint extension of the Hamiltonian is necessary and this will naturally lead to boundary conditions on the wave function.  Our main finding is that, in the context of a black hole, when boundary condition is coupled with first law of thermodynamics, it naturally gives area quantization. We show that the spacing is consistent with Bekenstein's old result. For GR it leads to Eq. (\ref{Bekenstein}).
   
   Thus, we make a direct quantization of entropy/area, unlike the earlier attempts, in the sense that the surface Hamiltonian (product of horizon entropy and temperature for a black hole) is quantized in the language of usual quantum mechanics. Therefore our method is completely new and gives a direct evidence of quantization of entropy. The most interesting outcome is that the evaluated result matches with that from the earlier ``indirect'' calculations. Hence we reconfirm Bekenstein's original value of Eq. (\ref{Bekenstein}), obtained by semi-classical approach.
   
   %We present our paper in the following order. Next section is devoted to summarize the earlier results on the surface Hamiltonian which lead to the main discussion of our paper. Section \ref{Quantization} is divided into three subsections. In the first subsection, we address the issues which make a difficult task to quantize Hamiltonian $xp$. Next an analysis is given to resolve the underlying problem. In the final subsection, the quantum of Hamiltonian is achieved. By using these, the quantization of horizon entropy of a black hole is done in Section \ref{Result}.  Finally we conclude in section \ref{Con}.

\section{\label{Hamiltonian}Surface Hamiltonian: a brief discussion}
   In this section, we shall briefly discuss about the surface part of the action of a gravitational theory so that a new reader can find the paper self-sufficient. Calculating this for a metric on the horizon, it will be shown that it has a thermodynamic interpretation. From there the surface Hamiltonian will be identified. Moreover, we shall discuss why such a Hamiltonian has $xp$ structure. Both the GR as well as more general theory like Lanczos-Lovelock gravity will be our attention. Here a summary of the required information, for clarity, will be introduced without any detailed calculation. An interested reader can discuss with the relevant references (e.g. \cite{Parattu:2013gwa,Chakraborty:2014joa}) for explicit analysis.  
   
   Very recently it has been shown that the thermodynamic structure of the gravitational theories can be discussed in terms of two variables. For GR these are given by $f^{ab}$ and $N^c_{ab}$ which are related to the usual variables by the following relations:
\begin{eqnarray}
f^{ab}&=&\sqrt{-g}g^{ab}~;
\nonumber
\\   
N^c_{ab}&=&-\Gamma^c_{ab}+\frac{1}{2}\Big(\Gamma^d_{ad}\delta^c_b+\Gamma^d_{bd}\delta^c_a\Big)~.
\label{fN}
\end{eqnarray}
Most notable point about these variables is that $N^c_{ab}$ is the conjugate momentum of $f^{ab}$.  Before going into the main idea, let us review some salient points. It is well known that, the Einstein-Hilbert (EH) Lagrangian {\it i.e.} $\sqrt{-g}R$ can be divided into two parts: one is quadratic in $\Gamma^a_{bc}$ and the other one is a total derivative part, we call them as bulk and surface parts respectively. Although these are not scalars individually, there are some important features associated with them. The bulk part alone gives Einstein's equations of motion without suffering any inconsistency as one does not need to impose vanishing of the both variation of metric and derivative of metric  at the boundary; here one just imposes the vanishing of $\delta g^{ab}$ at the boundary (See page $242$ of \cite{Paddybook}). On the other hand, the surface part is given by
\begin{equation}
\mathcal{L}_{\textrm{sur}} = \frac{2}{16\pi G}\partial_c\Big[Q_a^{~bcd}\Gamma^a_{bd}\Big]~,
\label{surface}
\end{equation}   
where $Q_a^{~bcd} = 1/2(\delta^c_ag^{bd}-\delta^d_ag^{bc})$. This term when calculated on the $r=\textrm{constant}$ surface for a static black hole, the action gives entropy in the near horizon limit. In this case the time integration is taken within the periodicity of Euclidean time (see page $663$ of \cite{Paddybook} for details). Of course, if one computes the total EH action for a spherically symmetric static metric it has a thermodynamical structure like $S-E/T$ where $E$ is identifies as the black hole energy \cite{Padmanabhan:2002sha}.
Moreover the surface part of the gravitational action can be expressed as
\begin{equation}
\mathcal{A}_{\textrm{sur}}=-\frac{1}{16\pi G}\int d^4x\partial_c\Big(f^{ab}N^c_{ab}\Big)~.
\label{Asur}
\end{equation} 
Therefore the surface Lagrangian has a structure like $\partial(xp)$ with $x\equiv f^{ab}$ and $p\equiv N^c_{ab}$.

   Now the calculation of this action on the null surface for static spacetime leads to
 \begin{equation}
\mathcal{A}_{\textrm{sur}}=-\frac{1}{16\pi G}\int dt d^2x_{\perp}n_cf^{ab}N^c_{ab}=-\int dt TS
\label{TS}
\end{equation}
where $n_c$ is the normal to the  surface, $x_\perp$ refers to the transverse coordinates and $T=\hbar\kappa/2\pi$ and $S=A/4G\hbar=A/4l_p^2$ are the horizon temperature and entropy, respectively with $\kappa$ being the surface gravity and $A$ is the area of the horizon. Therefore the surface Hamiltonian is identified as $H_{\textrm{sur}}=-\partial\mathcal{A}_{\textrm{sur}}/\partial t=(1/16\pi G)\int d^2x_{\perp}n_cf^{ab}N^c_{ab} = TS.${\footnote{It may be mentioned that the same $TS$ Hamiltonian can also be obtained from the Gibbons--Hawking--York surface term \cite{Majhi:2013jpk}, but it is not known if it has similar $xp$ structure. Therefore we restrict our discussion within the non-covariant form of the surface part of the main gravitational action.}} {\it From this we can immediately realise that the present Hamiltonian has $xp$ structure}. More precisely, Hamiltonian density (Hamiltonian per unit transverse area), which is temperature times entropy density (entropy per unit transverse area) has this structure. Moreover, among the two thermodynamical variables ($T$ and $S$), one of them plays the role of $x$ while the other one is $p$. Now to properly identify these variables we can take the help of the following analysis. It has also been observed that if we take a variation of the Hamiltonian; i.e. $\delta{H}_{\textrm{sur}}=1/16\pi G\Big[(\delta f^{ab})(n_cN^c_{ab})+(f^{ab})\delta(n_cN^c_{ab})\Big]$ and calculate them on the horizon, then these two parts lead to 
\begin{eqnarray}
&&\frac{1}{16\pi G}\int d^2x_{\perp}(\delta f^{ab})(n_cN^c_{ab})=T\delta S~;
\nonumber
\\
&&\frac{1}{16\pi G}\int d^2x_{\perp}(f^{ab})\delta(n_cN^c_{ab})= S\delta T~.
\label{delta}
\end{eqnarray}
Therefore one can say that $S\equiv x$ while $T$ is the conjugate momentum of $S$; i.e. $T \equiv p$. The details of these discussion can be followed from \cite{Parattu:2013gwa}. Now since the action is given by (\ref{TS}) and $T,S$ are conjugate variables, in classical mechanics we have the following Poisson's bracket:
 \begin{eqnarray}
 \{S,T\}_{\textrm{PB}}=1~.
 \end{eqnarray}  
 It should be mentioned that this feature is not restricted to GR; rather this is much more general. The same has also been concluded for a general Lanczos-Lovelock gravity.  For that we refer to \cite{Chakraborty:2014joa} for the readers.  In this general case the conjugate variables are $\tilde{f}^{ab}=f^{ab}$ and $\tilde{N}^c_{ab}=Q^{cd}_{ae}\Gamma^e_{bd} + Q^{cd}_{be}\Gamma^e_{ad}$ where $Q^{ab}_{cd}=(1/m) P^{ab}_{cd}$ with $P^{ab}_{cd}=\partial L_m/\partial R^{cd}_{ab}$ and $m$ is the order of the Lanczos-Lovelock Lagrangian $L_m$. The surface action is exactly in similar form with the un-tilde variables are replaced by tilde variables. Calculation of it on the horizon leads to identical results like the GR case with the entropy is properly defined in terms of the relevant component of $Q^{ab}_{cd}$.
{\it Hence we infer that the structure of surface Hamiltonian on the horizon is $TS$, which is more fundamental than in terms of temperature and horizon area}. On top of that, in a much more general theory of gravity, we have $S\equiv x$ and $T\equiv p$.

    After getting an interpretation of the surface Hamiltonian in terms of the usual $x$ and $p$, which are actually related to the thermodynamical quantities of a black hole, one would be curious to quantise it to see the quantum nature of black holes.  Therefore in the next section we shall discuss about the quantization of classical Hamiltonian $H=xp$. Before going into that let us mention about an earlier attempt to find this quantum nature using the boundary term of the action in GR \cite{Padmanabhan:2003qq,Padmanabhan:2003ub}.  It was shown that the in the context of ADM formalism the semiclassical path integral of gravity, after integrating out the unobserved degrees of freedom, depends on the boundary part of the action. Now imposition of the single valued nature of it leads to area quantization of a black hole. There also one can identify a Hamiltonian in terms of ADM variable and its conjugate momentum.  In the present discussion we are not using the ADM analysis. Instead we shall use here the information that the surface Hamiltonian, coming from the action without the Gibbons--Hawking--York (GHY) term (like Einstein-Hilbert action in GR),  has $xp$ structure in terms of $f^{ab}, N^c_{ab}$ ($\tilde{f}^{ab}, \tilde{N}^c_{ab}$ for more general theory). Just to mention that the ADM boundary part and the present one are not in general same.

\section{Quantization of $xp$ Hamiltonian }
  Though the Hamiltonian $xp$ looks simple, it has some unusual features. Note that, this Hamiltonian is of order 1 and it violates time reversal symmetry. The system described by this Hamiltonian has a hyperbolic point at $x=0, p=0$, and it represents the simplest form of instability.
Hamilton's equations of motion for this system are
\begin{equation}
\dot{x}=\{x,H\}_{PB}=\frac{\partial H}{\partial p}=x; \,\,\,\
\dot{p}=\{p,H\}_{PB}=-\frac{\partial H}{\partial x}=-p.
\label{classsolution}
\end{equation}
The solutions of the above equations are easy to find. These are given by
\begin{equation}
x(t)=x_0e^t; \,\,\ p(t)=p_0e^{-t}.
\label{xp}
\end{equation}
Thus classical solutions are dilation and contraction in $x$ and $p$ showing homogeneous instability. Now let us make an attempt to quantize this Hamiltonian. We follow the method described in \cite{Bonneau:1999zq}.

To begin with we take the normal ordered expression 
\begin{equation}
\hat{H}=\frac{1}{2}(\hat{x}\hat{p}+\hat{p}\hat{x})
\label{H}                                                                         
\end{equation}
as our Hamiltonian. It acts on the Hilbert space $\mathcal L^2(a,b)$. This Hamiltonian will be symmetric if
\begin{eqnarray}
 \braket{\psi|\hat{H}\phi}-\braket{\hat{H}\psi|\phi}=\int_a^b\psi^*\hat{H}\ph dx-\int_a^b\left(\hat{H} \psi\right)^*\ph dx=0.\label{hermitian}  
\end{eqnarray}
In the position space representation ($\hat{x}=x$ and $\hat{p}=-i\hbar\frac{d}{dx}$), we have
\begin{equation}
\hat{H}\ph =-\frac{i\hbar}{2}\left(x\frac{d\ph}{dx}+\frac{d}{dx}(x\ph)\right)=-\frac{i\hbar}{2}\left(2x\frac{d\ph}{dx}+\ph\right)
 \label{Hphi}                                                                         
\end{equation}
and
\begin{equation}
(\hat{H}\psi)^*=\frac{i\hbar}{2}\left(2x\frac{d\psi^*}{dx}+\psi^*\right).
 \label{Hpsi}                                                                         
\end{equation}
The above two expressions, when used in (\ref{hermitian}), gives 
$[x\psi^*\ph]_a^b=0$.
Note that $\hat{H}$ is not a self adjoint operator even if $\hat{H}^{\dagger}=\hat{H}$ because $\hat{H}^{\dagger}$ acts on different space. According to a theorem by von Neumann \cite{Bonneau:1999zq}, a symmetric operator like $\hat{H}$ is self-adjoint only if its deficiency indices are equal. To find the deficiency indices for $\hat{H}$, we have to solve the following equation
\begin{eqnarray}
\hat{H}^{\dagger}\psi_{\pm}(x)=\pm i \hbar\lambda\psi_{\pm}(x) \ ; \ \lambda> 0.
\end{eqnarray}
 Then deficiency indices ($n_+$ and $n_-$) are number of solutions for each sector. The solutions of the above equations are
\begin{eqnarray}
\psi_{\pm}(x)=\frac{C}{x^{\frac{1}{2}\pm\lambda}}.
\end{eqnarray}
This solution is meaningful only if we can find the normalization constant $C$ by using
\begin{eqnarray}
\braket{\psi_{\pm}|\psi_{\pm}}=\pm\frac{C^2}{2\lambda}\left(a^{\mp 2\lambda}-b^{\mp 2\lambda}\right)=1.
\end{eqnarray}
Now we can take different values of $a$ and $b$ and see whether self-adjoint extension of $\hat{H}$ is possible or not. Below we give a table (Table 1).\\
  \begin{table}[ht]
\begin{center}
\caption{Deficiency indices and self-adjoint extension of $\hat{H}$}

\label{table1}
~\\
\begin{tabular}{|c|c|c|}
\hline \hline
$(a,b)$ & $(n_+,n_-)$ & Whether $\hat{H}$ is Self-adjoint\\ \hline 
$a$= finite positive number, $b =\infty$ & $n_+=1,n_-=0$ & No \\ \hline
$a=0$, $b$=finite positive number& $n_+=0,n_-=1$ & No  \\ \hline
$a$ and $b$ are finite positive numbers, $b>a$ & $n_+=1,n_-=1$&Yes  \\ \hline \hline
\end{tabular}
\end{center}
\end{table}
Only the last case of this table is relevant for our purpose. According to von Neumann's theorem \cite{Bonneau:1999zq} when $n_+=n_->0$, there are infinite self-adjoint extensions. So for our case, self adjoint extensions of $\hat{H}$ are due to different boundary conditions parametrized by $U(1)$ i.e. by a phase $e^{i\theta}$. Thus the domain of $\hat{H}$ is
\begin{eqnarray}
\mathcal D(\hat H)=\{\psi,\hat{H}\psi\in {\mathcal L}^2(a,b);e^{i\theta}\psi(a)=\sqrt{\frac{b}{a}}\psi(b)\}\label{psi1}.
\end{eqnarray}
Now we consider the eigenvalue equation of (\ref{H}), 
$\hat{H}\psi =E\psi$. Since $\hat{H}\psi$ is given by (\ref{Hphi}), this will lead to
\begin{equation}
x\frac{d\psi}{dx}=(i\frac{E}{\hbar}-\frac{1}{2})\psi~.
 \label{HE}                                                                         
\end{equation}
 The solution of the above equation is
\begin{equation}
\psi(x)=\frac{C}{x^{\frac{1}{2}-\frac{iE}{\hbar}}}
 \label{solution}                                                                         
\end{equation}
where $C$ is the integration constant which can be fixed by the normalization condition. 
 The periodicity condition  (\ref{psi1}) of the wave function, when used for the above expression of eigenfunction, yields
$N^{iE/\hbar}=e^{i\theta}$ where $N=b/a$.
This immediately yields the energy quantization. The spectrum comes out to be 
\begin{equation}
E_n=\frac{2\pi\hbar}{\textrm{ln} N}(n+\frac{\theta}{2\pi})~,
 \label{En}                                                                         
\end{equation}
where $n$ is a natural number. This result and some mathematical applications of $xp$ Hamiltonian have been studied in \cite{Sierra:2007du}.

  Note that, in similar problems in physics, people usually take periodic boundary condition (i.e. $\theta=0$ in (\ref{psi1})). In that case energy (\ref{En}) turns out to be $\theta$ independent. Analysis presented in \cite{Bonneau:1999zq} shows that this choice of $\theta$, though simplest, is still arbitrary. In fact It has measurable consequences as explained in \cite{Bonneau:1999zq}. Thus only experiment can decide right value for it. However in the present study, $\theta$ appears as an additive constant in energy and wave function (\ref{solution}) does not depend on it ($\theta$ dependent eigen functions have been studied in \cite{Bonneau:1999zq}). Since we shall work with difference in energy levels and not with absolute energy, our results will be $\theta$ independent. 

 % It may be pointed out that the $xp$ Hamiltonian can also be quantized in an alternative way \cite{Paddyprivate}. The idea is to write it as a momentum like operator in a new variable. This is because we already discussed that the momentum operator can have quantum eigenvalues under a periodic condition on its eigenfunctions (See discussion around Eq. (\ref{periodic})). For that let us make the following transformation $x=e^{\mu}$ where $\mu$ is real. Then the operator form of the Hamiltonian transforms to $\hat{H} = -i\hbar x(\partial/\partial x) = -i\hbar(\partial/\partial\mu)$. See this is exactly identical to the momentum operator, represented in $\mu$ space here. Next using the Schr${\ddot{\textrm{o}}}$dinger equation $-i\hbar(\partial/\partial\mu)\psi=E\psi$, we obtain the eigenfunction as $\psi=\exp[iE\mu/\hbar]$. This has real eigenvalue provided $E$ is real. Next using the earlier hermiticity condition; i.e. Eq. (\ref{periodic}) and consequently following the arguments below Eq. (\ref{periodic}) we obtain the spectrum of the Hamiltonian as $E_n=\frac{2\pi\hbar n}{b-a}$.  Of course it would be interesting to investigate if these eigenvalues play any role in the quantum nature of black holes. For the present moment lets keep it as a side comment. Here we shall see below that the earlier result (i.e. Eq. (\ref{En})) is suitable for our present purpose.
  
  \section{\label{Result}Entropy quantization of a black Hole}
    Having obtained all the necessary inputs and equivalences, we are now going to discuss the quantization of the relevant thermodynamical quantity in the context of a black hole. Earlier it has been discussed that the surface Hamiltonian is like $xp$ Hamiltonian with $x\equiv S$ and $p\equiv T$. Quanization of this led to the spectrum, given by (\ref{En}). Here $a$ and $b$ were identified as the values of $x$ within which the Hamiltonian is self-adjoint. %where the wave function is periodic. 
    Now as we already mentioned, the surface action gives horizon entropy when one uses the periodicity of the Euclidean time; i.e. by Euclideanizing the surface action. Then the action has thermodynamical interpretation. So it is obvious that by doing Euclideanization, one obtains the black hole thermodynamics naturally and such a technique is quite normal in this paradigm (this can be understood by looking at the partition function).  Here our Hamiltonian is expressed in terms of the horizon temperature and entropy and all of them are meaningful in the sense of Euclidean signature. Also the present surface Hamiltonian is multiplication of two thermodynamical quantities and hence we take the evolving parameter as the Euclidean time. This we denote as $\tau$ here. This type of treatment is valid in the semi-classical regime. As the entropy plays the role of position here (this is a symbolic statement, not an exact one) and the classical solution is given by (\ref{classsolution}) with $x$ and $t$ are replaced by $S$ and $\tau$ respectively, we consider the interval $(a\equiv S_0,b\equiv S_0e^{\tau})$ for $x\equiv S$ within which the Hamiltonian $TS$ is self-adjoint. This is motivated by the fact that the Euclidean time is periodic within the interval ($0,\tau$) and then the solution (\ref{classsolution}) leads to this interval on $S$.   %to obtain a bounded solution we had to impose the condition that $S$ is bounded within the interval $(S_0,S_0e^{\tau})$. 
    %Now since entropy is a real quantity and it is associated with the horizon area, one must understand that the underlying input is the Euclideanization of the time coordinate. Therefore it is natural to choose $\tau$ as Euclidean time here.  
    %The classical solutions of equations of motions are presented by (\ref{xp}), which are unbounded. Since in the case of a black hole, the time coordinate is exactly the same time $t$, that appears in (\ref{xp}), the solutions can be bounded by using the periodicity of the Euclidean time of the metric.  Now suppose time is periodic within the interval ($0,\tau$), then using the equivalence between $(x,p)$ and ($S,T$), we can say $x\equiv S$ is bound within the interval $(S_0,S_0e^{\tau})$. 
    Thus the surface Hamiltonian $H_{{\textrm {sur}}}=TS$ has the quantum spectrum, given by (\ref{En}) with  $N=e^{\tau}$. Therefore, the spectrum of surface Hamiltonian for the black hole turns out to be
 \begin{equation}
 E_n=\frac{2\pi\hbar}{\tau}\left(n+\frac{\theta}{2\pi}\right)~.
 \end{equation}
 %where $\tau$ is the periodicity of the time coordinate of the metric. 
 
 %      Let us point out the logic behind the use of Euclidean time, instead of the coordinate time $t$. As we already mentioned, the surface action gives horizon entropy when one uses the periodicity of the Euclidean time. Then the action has thermodynamical interpretation. So it is obvious that by doing Euclideanization, one obtains the black hole thermodynamics naturally and such a technique is quite normal in this paradigm (this can be understood by looking at the partition function).  Here our Hamiltonian is expressed in terms of the horizon temperature and entropy. As entropy plays the role of position, to obtain a bounded solution we had to impose the condition that $S$ is bounded within the interval $(S_0,S_0e^{\tau})$. Now since entropy is a real quantity and it is associated with the horizon area, one must understand that the underlying input is the Euclideanization of the time coordinate. Therefore it is natural to choose $\tau$ as Euclidean time here.  
 
     Now the variation of (\ref{En}) due to change in energy level is given by
 \begin{equation}
 \Delta E=\frac{2\pi\hbar}{\tau}\Delta n .
 \label{delta_E}
 \end{equation}
 Let us assume $n$ is very large so that we can treat the variables of black hole physics as classical variables. In that approximation we can use the first law of thermodynamics: $ \Delta E=T \Delta S$. Then one obtains $\Delta S = \frac{2\pi\hbar}{T\tau}\Delta n$. As it has been already mentioned that all these are valid in the Euclidean signature, one can take  the periodicity of the Euclidean time as the inverse of the horizon temperature: $\tau=\hbar T^{-1}$. Therefore one finds that $\Delta S=2\pi\Delta n$ and hence the spacing between two consecutive levels is 
\begin{equation}
\Delta S=2\pi~,
\label{DeltaS}
\end{equation}
where $\Delta n=1$ has been used; i.e. the entropy is quantized. Note that this is true for any order Lanczos-Lovelock gravity as $H_{\textrm{sur}}=TS\equiv xp$ is not only restricted to GR. It may be mentioned that the fist law of thermodynamics, as used here, is meaningful in the sense of Euclideaniaztion. If one does not do that then both (\ref{delta_E}) and the first law of thermodynamics contain imaginary term consisting of the real time. But that will not change the final result as imaginary $i$ will get cancel in the intermediate step due to its presenc both in spectrum of energy and first law of thermodynamics. Moreover, all these calculations are valid in the semi-classical approximation. This type of semi-classical arguments has already been adopted earlier in \cite{Referee} which were restricted to the GR theory. 

   Since for GR, the entropy is proportional to area: $S=\frac{A}{4l_p^2}$, we can infer that the horizon area is also quantized and  the spacing between two consecutive levels is
\begin{equation}
\Delta A=8\pi l_{P}^2~. 
\end{equation}
This is Bekenstein's famous area quantization formula. It may be noted that the same was earlier re-obtained (after the original work by Bekenstein) by several methods, like using quasi-normal mode with the properly defined adiabatic invariant quantity \cite{Maggiore:2007nq}, tunnelling approach \cite{Banerjee:2009pf} -- \cite{Banerjee:2010be}, Euclideanization of time coordinate \cite{Ropotenko:2009mh} etc. Here we found that this result is consistent with a ``direct'' quantization of gravitational Hamiltonian on the horizon. Hence our approach is much more fundamental and does not rely on any outside input. Also remember that for other order Lanczos-Lovelock gravity, entropy-area relation is not so simple and therefore we do not have area quantization in general. Hence we may conclude that the entropy quantization is much more natural than that of area.

\section{\label{Con}Conclusions}
  One of the interesting facts of gravitational theories (GR as well as Lanczos-Lovelock gravity) is that the action can be written as a sum of two terms: one is the square of the Christoffel symbols and the other is the derivative of the Christoffel symbols. The latter is known as surface term ( See section $15.4$ of \cite{Paddybook}). The second term also has a thermodynamical interpretation. Calculating it on the horizon gives rise to a quantity which is horizon entropy times the temperature multiplied by the time coordinate. Consequently we can identify this as a surface Hamiltonian. One recent observation is that such a Hamiltonian can be interpreted as $xp$ for a spacetime with horizon \cite{Parattu:2013gwa,Chakraborty:2014joa}. This motivates us to quantize $xp$ Hamiltonian as this is directly related to the surface Hamiltonian for gravity. The idea behind this study is to get insight about the quantum nature of the horizon.
  
  One of the difficulties of quantizing this type of Hamiltonian lies in its non-hermitian nature. It has been found that one can avoid such burden by working with self-adjoint extension of the Hamiltonian. With this imposition the necessary condition on the wave function can be obtained which leads to the quantization. 
  
  In this paper we showed that this has something to do with the thermodynamics of gravity. One can talk about the quantization of horizon entropy. Here using the quantum spectrum of such Hamiltonian we found the quantum spectrum of entropy as well as surface area of a black hole event horizon. The obtained result exactly matches with Bekenstein's earlier finding \cite{Bekenstein:1973ur}. This provided a direct quantization in the sense that we used the quantum spectrum of the surface Hamiltonian which has been obtained by the standard procedure of quantum mechanics. Such an analysis is completely new, as the existing ones are based on several assumptions, like having to properly define the adiabatic invariant quantity \cite{Hod:1998vk,Majhi:2011gz}, time coordinate behaving like an azimuthal angle \cite{Ropotenko:2009mh} etc. Here we provided a more robust procedure to discuss the quantization of entropy without making any assumptions.
  
  We feel that our analysis may provide some deeper insights into the subject of area -- entropy quantization. Moreover, the procedure is general enough to be valid for any kind of black hole. Of course, it is evident from the main analysis that the quantum of entropy is much more fundamental than that of the area as in all cases we have to do entropy quantization first. Since for GR, entropy is proportional to area we obtain the area quantization; but where this is not the case, we can not talk about the quantization of area. Additionally, we note that the variables $f^{ab}$ and $N^c_{ab}$ ($\tilde{f}^{ab}$, $\tilde{N}^c_{ab}$ for Lanczos-Lovelock theory) have not only thermodynamic interpretation but also have structures which lead us to know more about the quantum nature of black hole parameters. Actually its identification with usual $x$ and $p$ gives us a way to think the gravitational theory along the path of the well established classical mechanics and corresponding quantum theory ({\it i.e.} quantum mechanics). Here we precisely driven by such logic and obtained the spectrum of the horizon entropy. Additionally, it may be worth to point out that here we are not quantizing the action directly as the Hamiltonian is identified by giving a thermodynamical interpretation of the surface action. The important observation is that the Hamiltonian has $xp$ structure in terms of thermodynamical entities and hence we are getting quantization of entropy.  Therefore, it has nothing to do with the quantization of action.
  
  Finally, let us mention that one aspect of our calculation may be interesting. Using the spectrum of $xp$ Hamiltonian, one can look at the several other thermodynamical quantities with the help of statistical mechanics procedure. We feel this will definitely enlighten several important aspects of gravity.  Another attempt can be more interesting. Considering the variables ($f^{ab}$ and $N^c_{ab}$) directly, without going into their thermodynamic interpretation, one may try to quantize the Hamiltonian $xp\equiv f^{ab}N^c_{ab}$. 
  It may be pointed out that the present analysis is based on the surface Hamiltonian which is $TS$. The same Hamiltonian (temperature times entropy) has also been obtained in \cite{Afshar:2016wfy} by multiplying the zero mode eigenvalue of the asymptotic charge, corresponding to the near horizon symmetry, by the surface gravity in three dimensions. Such an analysis exhibits that the modes of these charge satisfy infinitely many copies of the Heisenberg algebra. Exploiting these algebra one can find the Bekenstein-Hawking entropy \cite{Afshar:2016uax}. Since the analysis encodes the counting of microstates and the the central charge is essentially quantized, it may be possible to generate entropy quantization within this approach itself. Also it might be possible that our present analysis and near horizon symmetry approach may be interlinked as both of them uses the same Hamiltonian. Therefore we think it would be interesting to understand what is the algebra of symmetries that governs near horizon dynamics in dimensions other than three. Understanding this could imply and enhance many of the results we obtained here \cite{Grumiller}.    
  These will be studied in the coming future.
  
  \vskip 9mm
\section*{Acknowledgments}
We thank T. Padmanabhan and Daniel Grumiller for various interesting and valuable comments.
The research of one of the authors (BRM) is supported by a START-UP RESEARCH GRANT (No. SG/PHY/P/BRM/01) from Indian Institute of Technology Guwahati, India.

\end{document}